\documentclass[aps,prd,
showkeys,showpacs,
amssymb,preprintnumbers,nofootinbib,
superscriptaddress,twocolumn]{revtex4}

\usepackage{graphicx}
\usepackage{latexsym}
\usepackage{amsfonts}
\usepackage{url,hyperref}
\usepackage{bm}
\usepackage{amssymb}

\begin{document}

\begin{figure}[t]
\vspace{-1.4cm}
\hspace{-7.15cm}
\scalebox{0.085}{\includegraphics{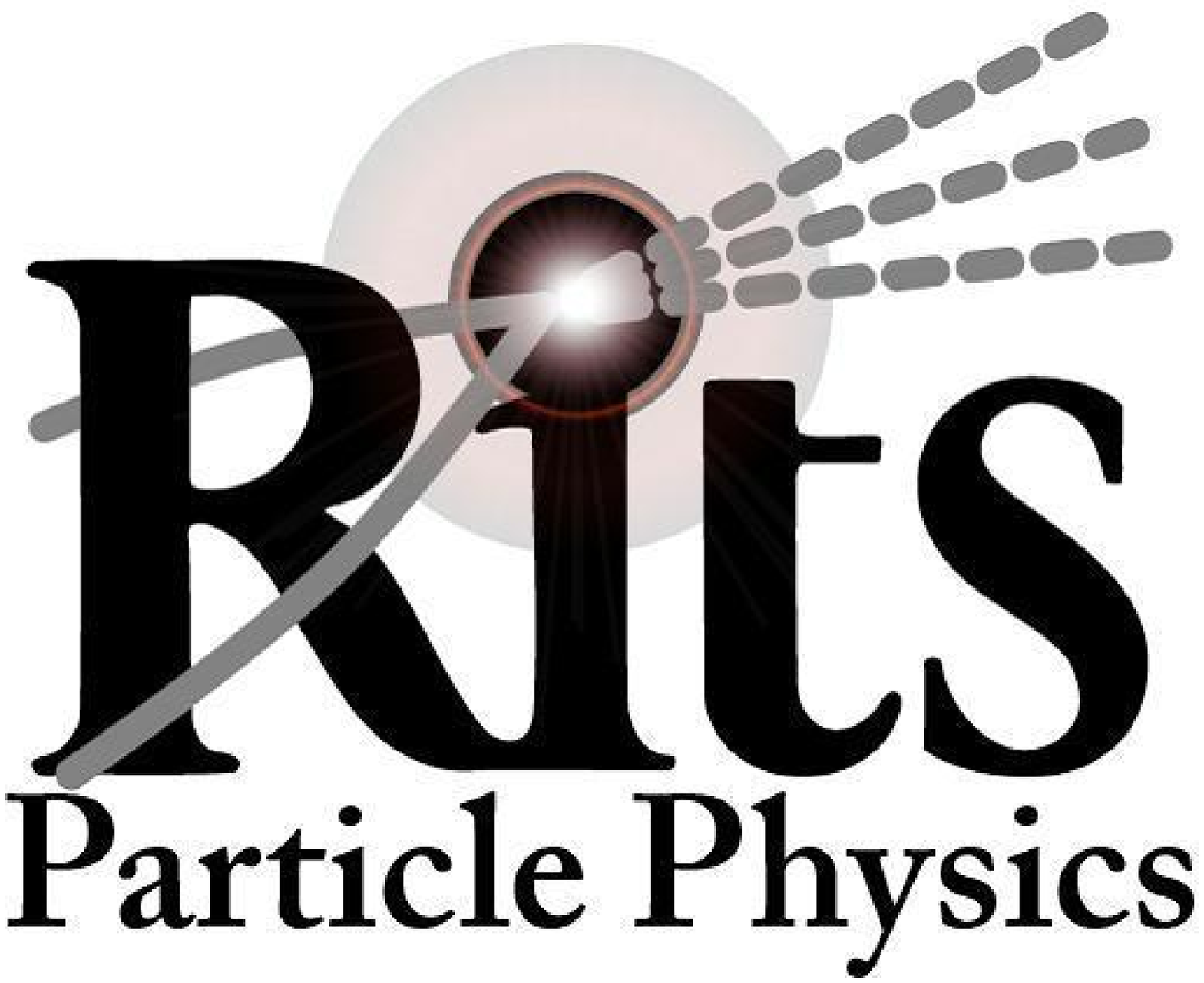}} 
\end{figure}

\newcommand{\vp}{\varphi}
\newcommand{\nn}{\nonumber\\}
\newcommand{\beq}{\begin{equation}}
\newcommand{\eeq}{\end{equation}}
\newcommand{\bed}{\begin{displaymath}}
\newcommand{\eed}{\end{displaymath}}
\def\bea{\begin{eqnarray}}
\def\eea{\end{eqnarray}}

%%%%%%%%%%%%%%%%%%%%%%%%%%%%%%%%%%%%%%%%%%%%%%%%%%%%%%
%\title{Quantum backreaction on a thick de Sitter brane}
\title{Can thick braneworlds be self-consistent?}
\author{Masato~Minamitsuji}
\email[Email: ]{masato@vega.ess.sci.osaka-u.ac.jp}
\affiliation{Department of Earth and Space Science, Graduate School of
Science, Osaka University, Toyonaka 560-0043, Japan}
\affiliation{Yukawa Institute for Theoretical Physics, Kyoto University, 
Kyoto 606-8502, Japan}
\author{Wade~Naylor}\email[Email: ]{naylor@se.ritsumei.ac.jp}
\affiliation{Department of Physics, Ritsumeikan University, 
Kusatsu, Shiga 525-8577, Japan}
\author{Misao~Sasaki}
\email[Email: ]{misao@yukawa.kyoto-u.ac.jp}
\affiliation{Yukawa Institute for Theoretical Physics, Kyoto University, 
Kyoto 606-8502, Japan}

\begin{abstract}
We discuss the backreaction of a massless, minimally coupled, quantized scalar field on a thick, two-dimensional de Sitter(dS) brane as an extension of our previous work \cite{Minamitsuji:2005gt}.
We show that a finite brane thickness {\it naturally} regularizes the backreaction on the brane. The quantum backreaction exhibits a quadratic divergence in the thin wall limit.
We also give a {\it theoretical} bound on the brane thickness, in terms of the {\it brane} self-consistency~of the quantum corrected Einstein equation, namely the requirement that the size of the backreaction should be smaller than that of the background stress-energy at the center of the  brane.
Finally, we discuss the {\it brane} self-consistency for the case of a four-dimensional dS brane.
\end{abstract} 

\pacs{04.50.+h; 98.80.Cq}
\keywords{Extra dimensions,~Quantum effects,~Cosmology}
\preprint{YITP-05-61}
\preprint{RITS-PP-007}
\date{\today}
\maketitle

%%%%%%%%%%%%%%%%%%%%%%%%%%%%%%%%%%%%%%%%%%%%%%%%%%%%%%%%%%%%%%%%%%%%%

Recent progress in string theory suggests that our 
universe is, in reality, a
four-dimensional submanifold,~{\it brane}, embedded 
into a higher-dimensional spacetime,~{\it bulk}.
The model which was proposed by Randall and Sundrum (RS)
succeeds in the localization of gravity around the brane due to the warping of the extra-dimension \cite{Randall:1999vf}.
This model has been given phenomenological grounds from various
aspects of higher-dimensional theories of gravity~\cite{Maartens:2003tw}.

In RS braneworlds, occasionally, fields which exist in the bulk affect the dynamics of the brane.
These bulk fields are naturally set into the bulk as a result of
a dimensional reduction of some higher-dimensional gravitational theory.
In some models such a bulk field plays the role of the inflaton, whose dynamics induces inflation on the brane  
(see references cited in \cite{Minamitsuji:2005gt}).
In these cases the bulk field carries information from the 
higher-dimensional {\it Kaluza-Klein} (KK) modes, as well as the zero mode, onto the brane.
Then, the problem we pose is whether or not the self-consistency of the quantum corrected Einstein equations can be kept on the brane:%%%
\footnote{
Self-consistency in the RS (two-brane) model has been investigated in \cite{Flachi:2001pq} and \cite{Knapman} in terms of how quantum corrections contribute to the gravitational theory in the bulk. Stability of brane solutions including quantum backreaction has also been discussed,~see e.g.,~\cite{Hofmann:2000cj}. What we do here is rather to compare the size of the quantum backreaction with that of the background stress-energy on the brane, i.e., {\it brane} self-consistency.}
% However, fully self-consistent solutions may require an analysis in the %bulk,% which is somewhat out of the scope of this letter.
The quantum backreaction of any bulk fields, especially for the KK modes, should be much smaller than the background stress-energy, such that the original features of the classical model are not changed.
Thus, the evaluation of the amount of backreaction is an important issue in these models.
The quantum backreaction for a thin de Sitter~(dS) brane has been considered in \cite{Pujolas:2004uj}, but little has been said about the self-consistency of the quantum backreaction on the brane.
Furthermore, it is well-known that for thin branes surface divergences remain on the brane even after UV regularization, which prevents one from evaluating any quantities exactly on the brane, e.g., see \cite{Knapman} and references therein. 

To overcome this pathology, in Ref.~\cite{Minamitsuji:2005gt} we investigated the quantum fluctuations for a thick dS brane based on the model discussed in \cite{Wang:2002pk} (for other thick brane models see the references mentioned in \cite{Minamitsuji:2005gt}). There, we demonstrated that the brane thickness can {\it naturally} regularize the surface divergences. In this letter we shall demonstrate that, similarly, a finite thickness also regularizes the quantum backreaction. %(i.e., the Hamiltonian density %and pressures).
We give a {\it theoretical} bound on the thickness in terms of {\it brane} self-consistency\footnote{Bounds on the brane thickness have also been discussed in terms of phenomenological experiments, see e.g.,~\cite{Casadio:2000pj}.} and comment on the realistic case of a four-dimensional brane.

The action of the system is given by
\begin{eqnarray}
S&=&\frac{1}{2}\int d^{d+1}x \sqrt{-g}
     \Bigl(
           \stackrel{(d+1)}{R}
          -\bigl(\partial \phi\bigr)^2
\nonumber\\ 
    &-&2V_0\bigl( \cos\bigl[\frac{\phi}{\phi_0}\Bigr] \bigr)^{2(1-\sigma)}
    -\bigl(\partial \chi\bigr)^2 \Bigr)
%\nonumber\\
%&+& \frac{1}{2}\int d^{d+1}x \sqrt{-g}
%      \Bigl(
%           -\bigl(\partial \chi\bigr)^2
%      \Bigr)\,,
     \label{action}
\end{eqnarray}
where the bulk field $\phi$, which is assumed to be static ($\phi=\phi(z)$), supports the thick brane configuration and $\chi(x^{\mu},z)$ is a test quantized scalar field propagating on the background produced by the scalar field $\phi$. Note that $-\infty<z<\infty$ represents the coordinate of the extra-dimension and we choose the center of the domain wall to be at $z=0$. We work with units such that the $(d+1)$-dimensional gravitational constant $\kappa_{d+1}^2=M_{d+1}^{-(d-1)}$ is set to unity, where $M_{d+1}$ is the $(d+1)$-dimensional Planck mass.  
Here, we are assuming that the field $\chi$ is a massless, minimally coupled scalar field, which is formally equivalent to tensor metric perturbations \cite{Minamitsuji:2005gt}. 
The thick dS brane solution is given by 
\begin{eqnarray}
ds^2&=&b^2(z)  \left(dz^2+\gamma_{\mu\nu}dx^{\mu}dx^{\nu}\right)\,;
\nonumber\\
&& b(z)=\left(\cosh\left(\frac{H z}{\sigma}\right)\right)^{-\sigma}\,,\quad
\nonumber\\
&& \phi(z)= \sqrt{(d-1)\sigma(1-\sigma)}
 \sin^{-1}\left(\tanh\left(\frac{H z}{\sigma} \right)\right)\,,
\nonumber\\
&& H^2= \frac{2\sigma V_0}{(d-1)[1+(d-1)\sigma]} \,;
\end{eqnarray} 
$\gamma_{\mu\nu}$ is the $d$-dimensional dS 
metric \cite{Wang:2002pk} with
$\gamma_{\mu\nu}dx^{\mu}dx^{\nu}=-dt^2+H^{-2}\cosh^2(Ht)d\varphi^2$ for $d=2$, where $-\infty< t< +\infty$ and $0\leq \varphi< 2\pi$. The brane thickness is parameterized by $\sigma$ (the physical thickness is $\sigma/H$), which is restricted to $0<\sigma<1$.

In this letter, we shall discuss the quantum backreaction of the scalar field $\chi$ on such a thick-brane background, specifically at $z=0$.
By varying the $\chi$-field part of the action, in Eq.~(\ref{action}), with respect to the bulk metric we obtain the stress-energy tensor for the $\chi$-field;
%\begin{eqnarray}
$T_{ab}:=\chi_{;a} \chi_{;b}
      -(1/2)g_{ab}
         g^{cd}\chi_{;c}\chi_{;d}\,$.
%\end{eqnarray}  

Furthermore, for simplicity we shall consider the three-dimensional ($d=2$) case. 
The method is then based on a dimensional reduction of the higher dimensional canonically quantized fields, see \cite{Naylor:2004ua}.
For a given vacuum, we can calculate the vacuum expectation value of the stress-energy tensor. Hereafter, we work in the Euclideanized space $\gamma^{\rm E}_{\mu\nu}dx^{\mu}dx^{\nu}=H^{-2}(d\theta^2+\sin^2\theta d\varphi^2)$,~where the substitution $\theta\to \pi/2- iH t$ Wick rotates back to the Lorentzian frame. Choosing the Euclidean vacuum  corresponds to a dS invariant vacuum in the original Lorentzian frame.  
The Hamiltonian density for the field $\chi$ in this frame is classically
defined by 
%\begin{eqnarray}
%&&
$ \rho(z,x^i):=
-b^2(z)T^{\theta}{}_{\theta}(z,x^i)\,$.
%\\
%&=&
%H^2\left(
%-\frac{1}{2}\left( \partial_{\theta} \chi\right)^2
%+\frac{1}{2}\frac{1}{\sin^2\theta}
%             \left( \partial_{\varphi} \chi\right)^2
%   \right)
%+\frac{1}{2}\left(\partial_z \chi\right)^2\nonumber\,.
%\end{eqnarray}

In general, for one non-trivial extra dimension we can have untwisted, $f^+(-z)=f^{+}(z)$, and twisted field configurations, $f^{-}(-z)=-f^{-}(z)$, \cite{Olum:2002ra, Graham:2002yr}.
Note that the untwisted and twisted solutions are equivalent to the mode degeneracy (for one non-trivial dimension). 
%For the case of the quantum fluctuations evaluated in \cite{Minamitsuji:2005gt}% the twisted field contribution vanishes at $z=0$; however, for the stress ener%gy tensor there is a non-zero contribution from the twisted solution at $z=0$.
As we shall see, the total Hamiltonian density is given by a  combination of untwisted and twisted fields, i.e., $\rho=\rho^{(+)}+\rho^{(-)}$.
This quantity diverges when all the modes are naively summed up and we need to employ some kind of regularization scheme. To this end, after a dimensional reduction, we shall employ the point-splitting method in conjunction with zeta function regularization~\cite{Minamitsuji:2005gt, Naylor:2004ua}, both for untwisted and twisted modes:
\begin{eqnarray}
\label{zetafunk}
&&\zeta^{\pm}(z,x^i,z',x^i{}';s)
:=\frac{2\mu^{2(s-1)}H}{b^{1/2}(z) b^{1/2}(z')H^{2(s-1)}}
\nonumber\\
&\times&\sum_{n} f^{\pm}_n(z)  f^{\pm}_n(z')
%\nonumber\\
%&\times&
\sum_{j,m} 
   \frac{Y_{jm}(x^i)Y^{\ast}_{jm}(x^{i}{}')}
          {\left[
               q_n^2  +(j+\frac{1}{2})^2 
           \right]^s}\,,
\end{eqnarray}
where $f_n^{+}(z)$ and $f_n^{-}(z)$ correspond to normalized untwisted and twisted field configurations respectively. The solutions $f_n^{\pm}(z)$ are written in terms of associated Legendre functions and $Y_{jm}(x^i)$ are the usual spherical harmonics defined on the two-sphere, $S^2$.

Given that we are interested only in the dependence of the backreaction on the bulk coordinate $z$ we integrate out over the trivial dimensions, in this case over $S^2$:
%\begin{eqnarray}
%\rho(z)&:=&
%\int  d\Omega_2 \rho(z,x^i)
%=\int  d\Omega_2
%\lim_{s\to 1}\lim_{X'\to X}
%\nonumber\\
%&\times&
%\frac{1}{2}
%\left(
%H^2
%\left(- \partial_{\theta} \partial_{\theta'}
%+\frac{1}{\sin^2\theta}
%            \partial_{\varphi}\partial_{\varphi'}
%\right)
%+\partial_{z}\partial_{z'}
%\right)
%\nonumber\\
%&\times&
% \zeta(z,x^i,z',x^i{}';s)
%\,,  \label{eucham}
%\end{eqnarray} 
\begin{eqnarray}
\rho(z)&:=&\int  d\Omega_2
\lim_{s\to 1}\lim_{X'\to X}
\nonumber\\
&\times&
\frac{1}{2}
\left(
  H^2 \partial_{\varphi}\partial_{\varphi'}
+\partial_{z}\partial_{z'}
\right)
% \tilde
 \zeta(z,x^i,z',x^{i}{}';s)
\end{eqnarray}
where $d\Omega_2$ is the volume element of $S^2$.
Note that because of the spherical symmetry transverse to the brane we may focus on the the equatorial plane $\theta=\pi/2$ and remove the dependence on $\theta$ \cite{Schwartz-Perlov:2005ds} Hence, we obtain the angle-integrated Hamiltonian density
\begin{eqnarray}
 \rho^\pm(z)
&=&2H%\frac{2\mu^{2(s-1)}H}{H^{2(s-1)}}
\lim_{s\to 1}
\sum_{n,j}\frac{j+\frac{1}{2}}
    {[q_n^2+(j+\frac{1}{2})^2 ]^s}\,\\
&\times&
\Bigg[
\frac{H^2}{2}
j(j+1)f_n^{\pm 2}(z)
+\Big(
 f_n^{\pm}{}'(z)
-\frac{b'(z)}{2b(z)} f_n^{\pm}(z)%f_n{}'(z)
 \Big)^2
\Biggr]
\nonumber\,.
\label{backreaction}
\end{eqnarray}

We are primarily interested in the backreaction on the brane at $z=0$, given that this is supposed to be where our world is localized.
In this case the contribution from the untwisted and twisted parts can be expressed simply as
\begin{eqnarray}
 &&
\rho^{+}(0)
=H\lim_{s\to 1}
%\frac{\mu^{2(s-1)}H}{H^{2(s-1)}}
\sum_{n,j}
H^2f_{n}^{+\,2}(0)
\frac{j(j+1)(j+\frac{1}{2})}
    {[q_n^2+(j+\frac{1}{2})^2 ]^s}\,,
\nonumber\\
&&\rho^{-}(0)
=2H\lim_{s\to 1}
%\frac{2\mu^{2(s-1)}H}{H^{2(s-1)}}
\sum_{n,j}
f^{-}_n{}^\prime{}^{2}(0)
%\Big(j+\frac{1}{2}\Big)
\frac{j+\frac{1}{2}}
    {[q_n^2+(j+\frac{1}{2})^2 ]^s}\,
 %   \nn
\end{eqnarray}
respectively. The total backreaction is given by the sum of each:
\begin{eqnarray}
\rho(0) 
=\rho^{+}(0)
+\rho^{-}(0)\,. \label{Hamilton}
\end{eqnarray}

We can also derive similar expressions for the pressures normal and parallel to the brane; $T^{z}{}_{z}(z,x^i)$ and $T^{\varphi}{}_{\varphi}(z,x^i)$, respectively.
Like for the Hamiltonian density they are given by a combination of untwisted and twisted fields.
In the Euclidean frame these pressures are defined by
%\begin{eqnarray}
%&&
$ p_z(z,x^i):= b^2(z) T^{z}{}_{z}(z,x^i)$
and
%\\
%&=&
%H^2\left(
%-\frac{1}{2} \left( \partial_{\theta} \chi\right)^2
%-\frac{1}{2}\frac{1}{\sin^2 \theta}
%             \left( \partial_{\varphi} \chi\right)^2
%\right)
%+\frac{1}{2}\left(\partial_z \chi\right)^2\,,
%\nonumber \\
%&&
$ p_{\varphi}(z,x^i)
:=b^2(z) T^{\varphi}{}_{\varphi}(z,x^i)\,$.
%\\
%&=&
%H^2\left(
%-\frac{1}{2} \left( \partial_{\theta} \chi\right)^2
%+\frac{1}{2}\frac{1}{\sin^2 \theta}
%             \left( \partial_{\varphi} \chi\right)^2
%\right)
%-\frac{1}{2}\left(\partial_z \chi\right)^2\,.
%\nonumber
%\end{eqnarray}
Then, as for the case of the Hamiltonian density, the angle-integrated total pressure normal to the brane $p_{z}(z)=\int  d\Omega_2 p_{z}(z,x^i)$ becomes 
\begin{eqnarray}
 p_z(0) =-\rho^{+}(0)+\rho^{-}(0)
\label{pressure}
\end{eqnarray}
on the brane. The angle-integrated pressure parallel to the brane ($p_{\varphi}(z)=\int 
d\Omega_2 p_{\varphi}(z,x^i)$) has the same amplitude with opposite sign, i.e., $p_{\varphi}(z)=-p_{z}(z)$. 
Thus, and hereafter, we may concentrate only on $p_{z}(0)$. 
It is also worth mentioning that the angle-integrated trace of the stress-energy tensor in the Euclidean frame has the same amplitude of the Hamiltonian density with the opposite sign;  
%\begin{eqnarray}
%&& b^2(z) T^a{}_a(z,x^i)
%\nonumber \\
%&=&-H^2\left(
% \frac{1}{2}\left(\partial_{\theta} \chi\right)^2
%+\frac{1}{2}
%\frac{1}{\sin^2\theta}  \left(\partial_{\varphi} \chi\right)^2
%\right)
%-\frac{1}{2}\left(\partial_z \chi\right)^2~.
%\nonumber\\
%\end{eqnarray}
%Hence, the angle-integrated trace is given by
%\begin{eqnarray}
$b^2(z)T^a{}_a(z)
=\int d\Omega_2 
b^2(z) T^a{}_a(z,x^i)
=-\rho(z)\,$.
%\end{eqnarray}
%Thus, once we know the result for the Hamiltonian density, all other related qu%antities can be trivially obtained. 

The regularization scheme we shall employ is basically the same as the one developed in Ref.~\cite{Minamitsuji:2005gt}. Essentially, we convert the mode sums over $\{ n \}$ into an integral along the contour as depicted in Fig.~4 of Ref. \cite{Minamitsuji:2005gt} by employing the residue theorem. We refer readers who are interested to the discussion in Sec. II.C of Ref. \cite{Minamitsuji:2005gt}.
For the twisted configuration there is no bound state and hence, we need not worry about how the contour approaches the imaginary axis. 

Hence, for untwisted configurations we obtain
\begin{widetext}
\begin{eqnarray}
\rho^{+}_{\rm UV}(0)
&=&-\frac{\sigma H^3}{2}
  \left[
     \int^{\infty}_{1}
       dU U^3  
    \left(
      \frac{P^{-U\sigma}_{\sigma/2}(0)}{P^{-U\sigma}_{\sigma/2}{}'(0)}
       -\sum_{\ell=0}^1 \frac{w_{\ell} (\sigma)}{U^{1+2\ell}}
   \right)
    -\sum_{\ell=0}^1 \frac{ w_{\ell} (\sigma)}{3-2\ell}
%\nonumber\\
%&+&
%+\frac{\sigma H^3}{8}
  -\frac{1}{4}
     \int^{\infty}_{1}
       dU U  
    \left(
      \frac{P^{-U\sigma}_{\sigma/2}(0)}{P^{-U\sigma}_{\sigma/2}{}'(0)}
       -\frac{w_0(\sigma)}{U}
   \right)
    - w_0(\sigma)
 \right]\,,
\nonumber\\
\rho^{+}_{\rm IR}(0)
&=&\frac{H^3}{2}
     \int_{0}^{1}
       dU  U  
  \left(U+\frac{1}{2}\right)
    \frac{\Gamma(-\frac{\sigma}{4}+\frac{U\sigma}{2}+1)
          \Gamma(\frac{\sigma}{4}+\frac{U\sigma}{2}+\frac{1}{2})}
         {\Gamma(\frac{\sigma}{4}+\frac{U\sigma}{2}+1)
          \Gamma(-\frac{\sigma}{4}+\frac{U\sigma}{2}+\frac{1}{2})}\,,
\end{eqnarray}
\end{widetext}
where we have split the untwisted contribution into two pieces, i.e., an ultraviolet (UV) and an infrared (IR) piece.
Here we used the following asymptotic expansion
%\begin{widetext}
\begin{eqnarray}
\frac{P^{-U\sigma}_{\sigma/2}(0)}{P^{-U\sigma}_{\sigma/2}{}'(0)}
&=&
\sum_{\ell=0}^{\infty} 
    w_{\ell}(\sigma)U^{-1-2\ell};
\nonumber\\
w_0(\sigma)&=&-\frac{1}{\sigma} ,~
w_1(\sigma)=-\frac{2+\sigma}{8\sigma^2} ,~
\cdots\,
%w_2(\sigma)
%=\frac{-16+4\sigma+12\sigma^2+3\sigma^3}
%      {128\sigma^4}~,
\end{eqnarray}
%\end{widetext}
to regularize the UV piece.
The regularized untwisted Hamiltonian density is given by
%\begin{eqnarray}
$\rho^+=\rho^+_{\rm UV}+\rho^+_{\rm IR}\,$.%\label{UVIR}
%\end{eqnarray}

Similarly, for twisted configurations we find
\begin{eqnarray}
\rho^{-}_{\rm UV}(0)
&=&\frac{H^3}{\sigma} \Biggl[
   \int_{1}^{\infty}dU\,
      U\Big(
  \frac{P^{-U\sigma}_{\sigma/2}{}'(0)}{P^{-U\sigma}_{\sigma/2}(0)}
\nonumber \\
&-&\sum_{\ell=0}^1 q_{\ell}(\sigma)U^{1-2\ell}
   \Big)
%\nonumber\\
% &-&
-\sum_{\ell=0}^1 \frac{q_{\ell}(\sigma)}{3-2\ell} 
   \Biggr]\,,
\nonumber\\
\rho^{-}_{\rm IR}(0)
&=&-\frac{2H^3}{\sigma}\int^{1}_0  dU  U
\nonumber\\
&\times&   
 \frac{\Gamma(\frac{\sigma}{4}+\frac{U\sigma}{2}+1)
          \Gamma(-\frac{\sigma}{4}+\frac{U\sigma}{2}+\frac{1}{2})}
         {\Gamma(-\frac{\sigma}{4}+\frac{U\sigma}{2})
          \Gamma(\frac{\sigma}{4}+\frac{U\sigma}{2}+\frac{1}{2})}\,,
\end{eqnarray}
where this time we used the slightly different asymptotic expansion  
\begin{eqnarray}
 \frac{P^{-U\sigma}_{\sigma/2}{}'(0)}{P^{-U\sigma}_{\sigma/2}(0)}
&=&\sum_{\ell=0}^{\infty} 
    q_{\ell}(\sigma)U^{1-2\ell};
\nonumber\\
q_0(\sigma)&=&-\sigma ,~
q_1(\sigma)=\frac{2+\sigma}{8} ,~
\cdots\,.
\end{eqnarray}
The twisted Hamiltonian density is also given by
%\begin{eqnarray}
$\rho^-= \rho^-_{\rm UV}
       +\rho^-_{\rm IR}\,$.
%\end{eqnarray} 

Before considering the total backreaction,  
we also wish to briefly discuss the backreaction for the bound state mode,~which for the untwisted case is given by
\begin{eqnarray}
\rho_{0}(z,s)
%&:=$
=
% -b^2(z) T^{+\theta}{}_{\theta,{\rm bs}}(z,s)
%=
\frac{H^3 \mu^{2(s-1)}f^{+}_{0}{}^2(z)}{H^{2(s-1)}}
%\nonumber\\
%&\times& 
\sum_{j}
\frac{j(j+1)(j+\frac{1}{2})}
   {[(j+\frac{1}{2})^2-(\frac{1}{2})^2 ]^s}~
\end{eqnarray}
where
\begin{eqnarray}
 f^{+}_0{}^2(z)
=\frac{1}{2\sigma}
\frac{ \cosh^{-\sigma}(\frac{Hz}{\sigma}) }
{\int^{\infty}_0 dy \cosh^{-\sigma}(y)}\,.
\end{eqnarray}
As can easily be verified the twisted solution has no localized bound state. Note that $\rho_{0}(z,s)=-p_{z,0}(z,s)$.
As a result of employing the renormalization discussed in \cite{Iellici:1997yh}, the backreaction for the bound state is given by
\begin{eqnarray}
\rho_0(z)
&=&
H^3
f_{0}^{+}{}^{2}(z)
\Bigl\{ 
 \zeta_{H}(-1,\frac{1}{2})
-\frac{1}{16} \zeta_{H}(3,\frac{1}{2})
+
\frac{1}{8}\\
&+& \sum_{J=2}^{\infty}  
   \Big(\frac{1}{2}\Big)^{2J}
\Big[
  \zeta_{H}(2J-1,\frac{1}{2})
-\frac{1}{4}\zeta_{H}(2J+1,\frac{1}{2})
\Big]
\Big\}\,.\nonumber
\label{bs}
\end{eqnarray}
In the case above (of massless, minimal coupling) the backreaction of the bound state mode is found to be independent of the renormalization scale. %%%%%%%%%%%%%%%%
%Although in general, if we consider a brane or bulk coupling to this field then% we do not expect the minimal case to be renormalization scale free. 
This happens only for the case of minimal coupling.
Also, contrary to the backreaction, 
%even without a brane or bulk coupling 
the squared amplitude itself does depend on the renormalization scale even for minimal coupling, $\xi=0$, \cite{Minamitsuji:2005gt}.
%%%%%%%%%%%%%%%%%%%%
If required, the contribution from the KK modes can easily be obtained  by employing the relations \cite{Minamitsuji:2005gt}
\begin{eqnarray}
\rho_{\rm KK}(z)=  \rho(z)-\rho_0(z)\,,\quad
p_{\rm KK}(z)=  p(z)-p_0(z)\,.
\end{eqnarray}
Note that, like for the amplitude \cite{Minamitsuji:2005gt}, $\rho_0$ is also insensitive to the brane thickness, $\sigma$.  Particularly in comparison to the KK contribution. 
 
Similar to the calculation for the amplitude \cite{Minamitsuji:2005gt} in the thin wall limit, $\sigma\to 0$, the leading order behavior for $\rho^{\pm}(0)$ can easily be obtained
\begin{eqnarray}
&&\rho^+(0)\to -\frac{H^3}{2 \sigma^2} I \,,\quad
 \rho^-(0)\to -\frac{H^3}{\sigma^2} I ;
\nonumber\\
 I&:=&\frac{1}{4}\int^{\infty}_{0}
  dx \Bigl[
     x^2  \left(
          \psi(\frac{x}{2})
         -2\psi(\frac{x+1}{2})
         +\psi(\frac{x}{2}+1)
        \right)
      +1  
     \Bigr]
\nonumber \\
&  \thickapprox &
    0.213139   \label{thin}
      \,.
\end{eqnarray} 
Thus, from Eq.~(\ref{Hamilton}) and Eq.~(\ref{pressure}), the thin wall behavior for the Hamiltonian density and pressure are
\begin{eqnarray}
\rho(0)\to -\frac{3H^3}{2\sigma^2} I \,,
\quad
p_z(0)\to-\frac{H^3}{2\sigma^2} I\,,
\end{eqnarray}  
respectively. Thus, in the thin wall limit both the total Hamiltonian density and pressure exhibit quadratic divergences. 
%This is one of the main results in this letter.

We now come to discuss the {\it brane} self-consistency 
of the quantum corrected Einstein equations: The stress-energy of the backreaction should not become larger than the background stress-energy on the brane (see footnote 1).  
In Figs.~1 and 2 the Hamiltonian density, $\rho(0)$, and bulk pressure, $p_{z}(0)$, are compared with respect to their (angle-integrated) classical counterparts:
\begin{eqnarray}
%&&\int d\Omega_2
%  b^2(z)T^{\alpha}{}_{\beta}(z,x^{\mu})
%\\
%&=&
&&
4\pi\left( -\frac{1}{2}\left(\phi'(z) \right)^2-b^2(z)V(\phi(z))\right)
\delta^{\alpha}{}_{\beta}
\nonumber\\
&=&
-\frac{4\pi H^2}{\sigma\cosh^2\left(\frac{Hz}{\sigma}\right)}
\delta^{\alpha}{}_{\beta}
\stackrel{z=0}{\to}
-\frac{4\pi H^2}{\sigma}\delta^{\alpha}{}_{\beta}\,,
\nonumber \\
%&&\int d\Omega_2
%  b^2(z)T^{z}{}_z(z,x^{\mu}) 
%\\
%&=&
&&4\pi\left( \frac{1}{2}\left(\phi'(z) \right)^2-b^2(z)V(\phi(z))\right)
\nonumber\\
&=&-\frac{4\pi H^2}{\cosh^2\left(\frac{Hz}{\sigma}\right)}
 \stackrel{z=0}{\to}
-4\pi H^2\,\nonumber
\end{eqnarray}
for the special case $H=1=M_3$.
Note, that in Fig. 1 the Hamiltonian density is multiplied by a power of $\sigma$, in order to easily distinguish between the two,~and the three-dimensional Planck mass $M_3:=\kappa_3^{-2}$ is set to unity.
   
The quantum backreaction scales as $H^3/\sigma^2$, whereas the background stress-energy scales as $H^2 M_3 /\sigma$.
Thus, the ratio of the backreaction to the background energy density scales as $O(H/M_3\sigma)$. 
From Figs. 1 and 2, for the special case $H=M_3$, we can infer that for brane thicknesses with $\sigma \gtrsim  0.3$ the quantum backreaction is at least an order of magnitude smaller than the classical value.
Thus, taking these facts into consideration, we obtain a plausible theoretical bound on the brane thickness, $\sigma \gtrsim 0.3 \,(H/M_3)$. 
Of course, this bound is only valid on the brane not in the whole bulk. In this sense it might not be a sufficient condition, but just a necessary one.
However, we are mainly interested in the behavior of the quantum backreaction at $z=0$, where the backreaction is expected to be largest and our world exists by assumption. Thus, it may be considered as a stringent bound on the brane thickness.

%%%%%%%%%%%%%%%%%%%%%%%Fig1&2%%%%%%%%%%%%%%%%%%%%
\begin{figure}[htbt]
\begin{center}
  \begin{minipage}[t]{.45\textwidth}
 \begin{center}
  \includegraphics[scale=.75]{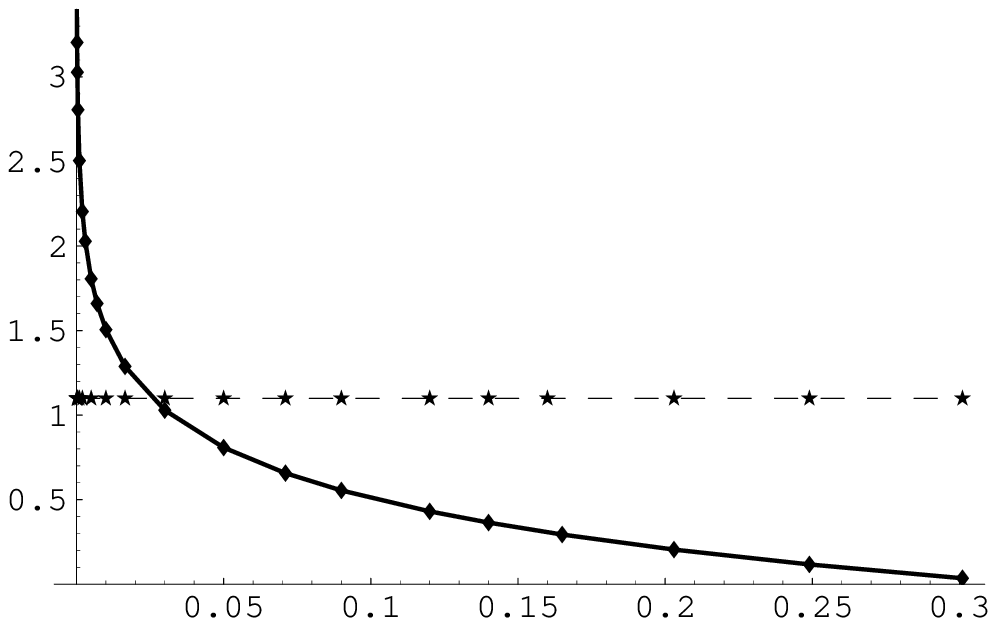}
\caption{
The backreaction of the Hamiltonian density (thick curve) and the background energy density (dashed curve), multiplied by a power of the brane thickness, $\sigma$, are shown as a function of $\sigma$ for the case of $H=1(=M_3)$.
Note, the scale of the vertical axis is set to $\log_{10}$. 
} 
  \end{center}
\end{minipage}
\begin{minipage}[t]{.45\textwidth}
\vspace{0.5cm}
\begin{center}
  \includegraphics[scale=.75]{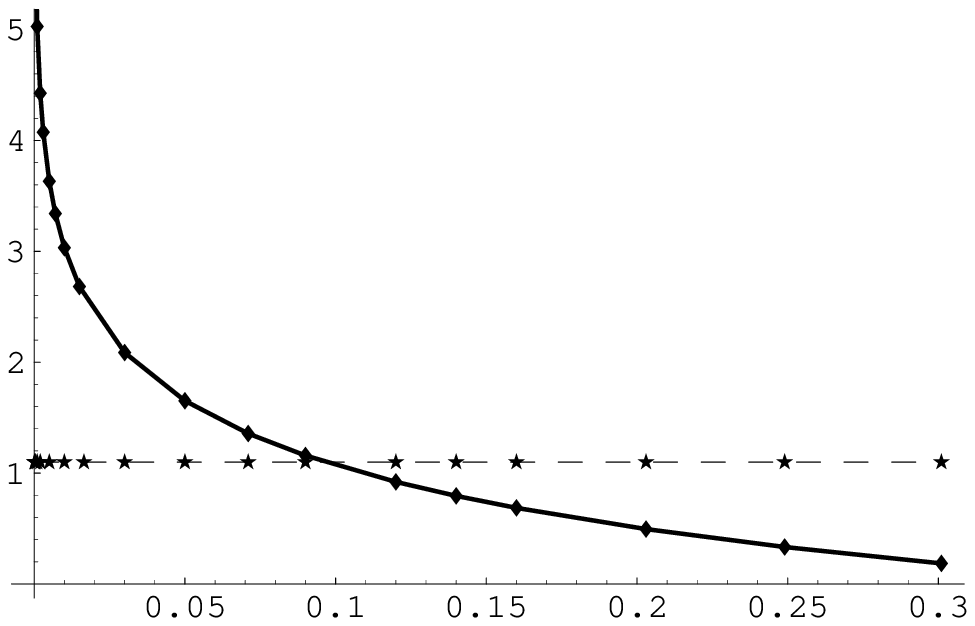}
\caption{
The backreaction of the pressure (thick curve) and the background pressure (dashed curve) are shown as a function of 
the brane thickness, $\sigma$, for the case of $H=1(=M_3)$.
Note, the scale of the vertical axis is set to $\log_{10}$. 
}
\end{center}
 \end{minipage}
 \end{center}
\end{figure}
%%%%%%%%%%%%%%%%%%%%%%%%%%%%%%%%%%%%%%%%%%%%%%%%%%%%%%%%%%

Now let us consider the more realistic case of $d=4$.
Following steps similar to that in (\ref{thin}), it is not hard to convince ourselves that in the thin wall limit, $\sigma\to 0$, the quantum backreaction exhibits a quartic divergence proportional to $H^5/\sigma^4$ for $d=4$ (times a numerical factor, like $I$ in Eq.~(\ref{thin}) for $d=2$), whereas the background stress-energy scales as $H^2 M_5{}^3/\sigma$.
Thus, the ratio of the quantum backreaction to the background stress-energy will be of order $O( H^3/(\sigma M_5)^3)$. 
Therefore, the brane should satisfy $\sigma \gtrsim H/M_5$ in order to have {\it brane} self-consistency. %, which is exactly parallel to the $d=2$ case.
Actually, this bound depends on the ratio between the energy scale for brane inflation, $H$, and $M_5$.
Hence, in order for the bound to be consistent with the assumption 
$0<\sigma<1$ we should require $H \lesssim  M_5$. This condition can also be regarded as a bound on the energy scale for brane inflation.

We might ask whether or not this bound on the brane thickness is consistent within the framework of the background model.
The thick brane model we have investigated in this letter has an asymptotically flat bulk, which can be regarded as the
high-energy limit ($H\ell\to \infty$) of an asymptotically Anti de Sitter (AdS)
bulk,
where $\ell$ is curvature radius of AdS spacetime.
Quite clearly $\sigma \ll H\ell$ and thus, combining this inequality with the previous theoretical bound,
$\sigma \gtrsim H/M_5$, we find that $H/M_5  \ll H\ell$
for {\it brane} self-consistency. Note that in the RS II set-up the four-dimensional Planck scale on the brane effectively becomes 
$M_{\rm pl}{}^2=M_{5}{}^3\ell\,(\approx 10^{19} {\rm GeV})$, which is determined at low 
energies ($H \ell \ll 1$) \cite{Randall:1999vf,Garriga:1999bq}.
Then, {\it brane} self-consistency, $H/M_5 \ll H\ell$, is equivalent to the condition $M_{\rm pl} \gg M_5$, which seems to be quite a natural one.  
Indeed, it is not difficult to construct a model with $M_5\ll M_{\rm pl}$; just as long as the scale is larger than
$10^9~{\rm GeV}$, %(the bound $10^9~{\rm GeV}$is 
derived from constraints on the size of any extra-dimensions, $\ell \lesssim 0.1{\rm mm}$, which is determined from experimental tests of Newton's law on short distance scales.
Thus, we conclude that thick braneworlds, even if they are extremely thin, can be {\it brane} self-consistent.

Finally, we shall comment on the quantum backreaction of the KK gravitons on the brane, which are considered to be produced during brane inflaton, see e.g., \cite{Langlois:2000ns}. The quantification of the graviton backreaction has been a longstanding issue in brane 
cosmology. (The classical nature of the backreaction for the KK gravitons on the brane, in the thin wall approximation, was partially investigated in \cite{Minamitsuji:2005xs}.) 
Part of the motivation in this letter 
%for studying a massless, minimally coupled scalar field 
is that the KK gravitons satisfy the same equation of motion as a massless, minimally coupled scalar field and therefore suffer from a similar pathology on the brane, {\it surface divergences}. 
Hence, we expect that the graviton backreaction behaves in a similarly manner to the scalar case, namely the backreaction exhibits quadratic and quartic divergences for $d=2$ and $d=4$, respectively.
Furthermore, any discussion on the self-consistency of the scalar backreaction should also carry over to the graviton backreaction similarly, though any explicit demonstration of this fact is left for future work.
 
In conclusion, in this letter we have explicitly discussed the backreaction of a quantized bulk scalar field on a two-dimensional
thick de Sitter brane.
As expected, a finite thickness {\it naturally} regularizes the surface divergences arising from the KK modes. We also obtained a {\it theoretical} bound on the size of the brane thickness in compliance with the {\it brane} self-consistency of the backreaction and then discussed the case of a four-dimensional brane.
The investigation of the full {\it bulk} self-consistency is left for another time. Also, as an extension, it would be interesting to investigate the {\it local} quantum effects on a cosmological brane in a higher-codimensional bulk.~% (see \cite{Saharian} for the thin-brane case). 
We hope to report on these issues in future publications.

We are grateful to A. Knapman for a critical reading of this manuscript. This work was supported in part by Monbukagakusho Grant-in-Aid for Scientific Research(S) No. 14102004 and (B) No.~17340075.

%%%%%%%%%%%%%%%%%%%%%%%%%%%%%%%%%%%%%%%%%%%%%%%%%%%%%%%%%%%%%%

%%%%%%%%%%%%%%%%%%%%%%%%%%%%%%%%%%%%%%%%%%%%%%%%%%%%%%%%%

\end{document}